\documentclass[11pt,oneside]{article}

\usepackage{a4wide}

\usepackage{amsmath}
\usepackage{color}
\usepackage{framed}
\usepackage{tikz}
\usepackage{tikz-cd}

\RequirePackage{amsmath}
\RequirePackage{amssymb}
\RequirePackage{amsthm}
\RequirePackage{color}
\RequirePackage{url}
\RequirePackage{mdwlist}

\RequirePackage{rotating}

\RequirePackage[all]{xy}
\RequirePackage{graphicx}

\usepackage{xcolor}
\usepackage{amsmath,amsfonts,amssymb}
\usepackage{graphicx}
\usepackage[caption=false]{subfig}
\usepackage{enumerate}
\usepackage{mathrsfs}
\usepackage{mathtools}

\makeatletter
\newcommand{\verbatimfont}[1]{\renewcommand{\verbatim@font}{\ttfamily#1}}
\makeatother

\newcommand{\Natural}{\mathbb{N}}

\usepackage{mathtools}

\newcommand\nounderline[1]{ #1} 
\newcommand\dolemma[1]{\vskip 5pt \noindent{\bf \underline{Lemma #1.}\ }}

\newcommand\dotheorem[1]{\vskip 5pt \noindent {\bf \underline{Theorem #1.}\ }}

\newcommand\dodefn[1]{\vskip 5pt \noindent {\bf \underline{Definition #1.}\ }}
\newcommand\docorollary[1]{\vskip 5pt \noindent {\bf \underline{Corollary #1.}\ }}

\newcommand\doproof{\vskip 5pt \noindent{\bf \nounderline{Proof:}\ }}

\newcommand\tombstone{\rule{.6em}{.6em}}

\newcounter{numitem}

\title{Entropy, Derivation Operators and Huffman Trees}

\author{Simon Burton\\
{\small{\it Institute of Physics, Jagiellonian University, \L{}ojasiewicza 11, 30-348 Krak\'ow, Poland}}}

\date{\today}

\begin{document}

\maketitle

\abstract{
We build a theory of binary trees on finite multisets
that categorifies, or operationalizes, the entropy of a finite
probability distribution. 
Multisets operationalize probabilities as the event outcomes of an experiment.
Huffman trees operationalize the entropy of the distribution of these events.
We show how the derivation property of the entropy of a joint distribution
lifts to Huffman trees.
}

%
%
%

\section{Entropy is...}

We define an {\it un-normalized} probability distribution $x$
of length $n$ to be a sequence of non-negative real values
$$
    x = (x_1,...,x_n).
$$
The {\it norm} of an un-normalized probability distribution $x$ is
the sum of the components:
$$
    |x| = \sum_{i=1}^n x_i.
$$

We look for a continuous real valued function $w$ on a variable number of arguments
such that
\begin{align}
w(x_{11}, \ldots, x_{1 k_1}, \ldots, x_{n 1}, \ldots, x_{n k_n})
=
w\biggl(\sum_{j = 1}^{k_1} x_{1 j}, \ldots, \sum_{j = 1}^{k_n} x_{n j}\biggr)
+
\sum_{i = 1}^n w(x_{i 1}, \ldots, x_{i k_i})
\end{align}
for real values $c, x_i\ge 0,$ and
\begin{align}
w(c x_1, \ldots, c x_n) = c w(x_1, \ldots, x_n)
\end{align}
for real values $c, x_i\ge 0.$
Up to scale this has unique solution
$$
w(x_1, \dots, x_n) = \sum_{i=1}^n x_i \log_q x_i 
    - \biggl(\sum_{i=1}^n x_i\biggr) \log_q \biggl(\sum_{i=1}^n x_i\biggr),
$$
with the convention that $a\log a$ is zero when $a$ is zero.
The scale corresponds to choice of logarithm base $q.$

We consider this sequence of non-negative reals $(x_1,...,x_n)$ to
be an \emph{un-normalized probability distribution,}
and the function $w(\cdot)$ is the \emph{un-normalized entropy.}
When we have $\sum_i x_i=1$ then
$$
    -w(x_1,...,x_n) = -\sum_{i=1}^n x_i \log x_i = h(x_1,...,x_n)
$$
where $h$ is the usual entropy function of the probability
distribution $(x_1,...,x_n).$

Given two un-normalized probability distributions
$x = (x_1,...,x_m)$ and $y=(y_1,...,y_n)$
we define the product as the length $mn$ un-normalized probability
distribution:
$$
    xy := (x_1 y_1, ..., x_1 y_n, x_2 y_1, ..., x_m y_n).
$$
Then we have that $w(\cdot)$ is a {\it derivation:}
\begin{align}
    w(xy) = |x| w(y) + w(x) |y|.
\end{align}
This follows from the fact that the function
$f(a) = a\log a,$ satisfies $f(ab) = f(a)b + af(b).$

Our aim below will be to categorify, or operationalize,
the equations (1), (2) and (3).

%
%
%

\section{Multisets categorify probabilities}

The first step is to replace real numbers by the natural numbers $\Natural$,
following the work of~\cite{Fritz2019}.
Given a finite set $A,$
a {\it multiset} over $A$ will be the set of $\Natural$-linear
formal combinations of elements of $A$.
We denote the set of multisets over $A$ as $\Natural[A]$.
We think of $A$ as a space of events and a
multiset $X\in \Natural[A]$ as
a record of how many times we have observed each event.

We either write multisets using set notation $X = \{a, b, 2c\}$
or as a formal sum $X = a+b+2c.$
The {\it zero multiset}, $0\in\Natural[A]$ has all zero coefficients,
$0 = \sum_{a\in A} 0\cdot a.$

The {\it norm} of
a multiset $X\in\Natural[A]$ is denoted $|X|\in\Natural.$
For $X = \sum_{a\in A} n_a\cdot a$ we have $|X| := \sum_{a\in A} n_a.$
This is the total number of observations.

Elements of $\Natural$ are called {\it scalars.}
For $k\in \Natural$ we write $kX$ for the left multiplication of $k$ on $X.$
This is given by $kX := \sum_{a\in A} (k n_a)\cdot a.$
We also call $kX$ a {\it scalar multiple} of $X.$

If we are also given $Y\in \Natural[B]$ as $Y = \sum_{b\in B} m_b\cdot b,$
the sum $X+Y\in\Natural[A\cup B]$ is
$$
    X+Y := \sum_{a\in A\cup B} (n_a + m_b)\cdot a.
$$
The product $X\times Y\in\Natural[A\times B]$ is
$$
    X\times Y := \sum_{(a,b)\in A\times B} (n_a m_b)\cdot ab.
$$
where we understand $ab$ to be shorthand notation for the pair $(a,b).$
These are the usual rules of polynomial addition and multiplication.
A {\it monomial} is a multiset with just one non-zero component,
such as $x \cdot a\in\Natural[A]$ with $x\in\Natural, a\in A.$

The intersection $X\cap Y\in\Natural[A\cap B]$ is
$$
    X\cap Y := \sum_{a\in A\cap B} (n_a m_a)\cdot a.
$$
Two multisets $X$ and $Y$ are {\it disjoint} when $X\cap Y=0.$

We extend the definition of the un-normalized entropy function $w(\cdot)$ to multisets.
Given  $X = \sum_{a\in A} n_a\cdot a$, we define
$$
    w(X) := w(\{n_a\}).
$$
This is well defined because the function $w(\cdot)$ is commutative in
its arguments.

%
%
%
%
%
%
%
%
%

%
%
%

\section{Trees over multisets}

\newcommand{\Tree}{\mathrm{Tree}}
\newcommand{\Leaf}{\mathrm{Leaf}}
\newcommand{\Depth}{\mathrm{Depth}}

The next definition formalizes a notion of binary trees with 
nodes decorated by multisets.
Here we stick to binary trees, even though this theory
generalizes to a theory of $q$-ary trees with $q=2,3,4,...$

\dodefn{1}
A {\it tree over a multiset} is defined recursively as\\
(1) For any multiset $X$, $\Tree(X)$ is a tree over $X$.\\
(2) If $X$ and $Y$ are disjoint multisets,
and $\Delta_X$ is a tree over $X$ and
and $\Delta_Y$ is a tree over $Y$, then
$\Tree(X+Y, \Delta_X, \Delta_Y)$ is a tree over $X+Y.$
\tombstone

\vspace{5pt}
Notation such as $\Delta_X$ will denote a
generic tree over the multiset $X$.
We consider $\Tree(\cdot,\cdot,\cdot)$ to be commutative in the last two arguments, so that
$\Tree(X+Y, \Delta_X, \Delta_Y)$ and 
$\Tree(X+Y, \Delta_Y, \Delta_X)$ denote the same tree.

Informally,
a tree corresponds to a way of parenthesizing the expression for $X$.
For example, given $X=a+b+2c\in\Natural[\{a,b,c\}]$ the tree 
$$
\Tree(a+b+2c, \Tree(a+b, \Tree(a), \Tree(b)), \Tree(2c))
$$
corresponds to the parenthesized expression $(((a)+(b))+(2c)).$
We also use a graphical notation:
\begin{center}
\includegraphics[scale=0.9]{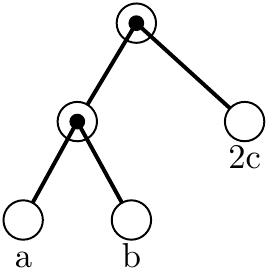}
\end{center}

Scalars and multisets act on trees in the same way, which
we formalize in the following two definitions.
\dodefn{2}
Given a scalar $k\in\Natural,$ and multisets $X, Y$,
we have that $k$ acts on trees recursively as\\
(1) $k\cdot\Tree(X) := \Tree(k\cdot X)$.\\
(2) $k\cdot\Tree(X+Y, \Delta_X, \Delta_Y) := \Tree(k\cdot(X+Y), k\cdot\Delta_X, k\cdot\Delta_Y).$
\tombstone
\vspace{5pt}

\dodefn{3}
A multiset $X$ acts on trees recursively as\\
(1) $X\cdot\Tree(Y) := \Tree(X\times Y)$.\\
(2) $X\cdot\Tree(Y+Z, \Delta_Y, \Delta_Z) := \Tree(X\times(Y+Z), X\cdot\Delta_Y, X\cdot\Delta_Z).$
\tombstone
\vspace{5pt}

We think of this action as {\it $X$-thickening} a tree.
For example, the multiset $a+b$ acting on the above tree $(((a)+(b))+(2c))$ is
\begin{center}
\includegraphics[scale=0.9]{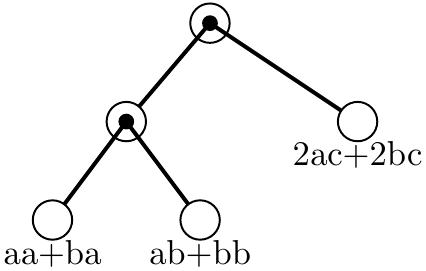}
\end{center}

\dodefn{4} For $X$ a multiset, the predicate {\it $X$ is a leaf of $\cdot$}  
is defined recursively as\\
(1) $X$ is a leaf of $\Tree(X).$ \\
(2) If $X$ is a leaf of $\Delta_Y$ then $X$ is a leaf
of any tree $\Tree(Y+Z, \Delta_Y, \Delta_Z).$
\tombstone
\vspace{5pt}

When $X$ is a leaf of the tree $\Delta_Y$ 
we also write this predicate as $\Leaf(X,\Delta_Y).$
From Definition 1, we see that if a multiset $Y$ is a leaf
of a tree $\Delta_X$ then this leaf is the unique occurrence of $Y$ in $\Delta_X.$
In this case, we define the {\it sum of trees} $\Delta_X+\Delta_Y$
by attaching $\Delta_Y$ at this leaf.
Instead of formalizing this with another recursive definition,
we show an example:
\begin{center}
\includegraphics[scale=0.9]{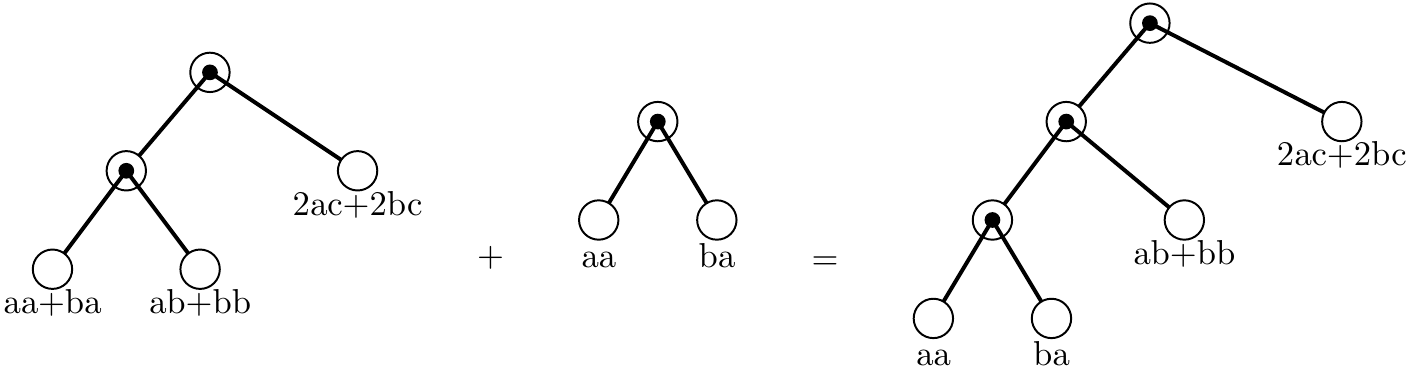}
\end{center}

The {\it product} of two trees $\Delta_X$ and $\Delta_Y$ is
a sum of trees, which we write as
$$
    \Delta_X\times\Delta_Y = X\Delta_Y + \Delta_X Y.
$$
This is not quite a formal definition of the product, but
it is close. We do have a definition of $X\Delta_Y$ as
an $X$-thickened $\Delta_Y.$ We could also try to formalize
$\Delta_X Y$ as a $\Delta_X$-thickened multiset 
and then hope that $X\Delta_Y + \Delta_X Y$ really is a sum of all
these trees, but this only works up to some judicious applications of associativity. 
We write this formula for the product because it suggests
that ``tree'' is a derivation over multisets.

Here we show another example, that builds on the previous examples:

\begin{center}
\includegraphics[scale=0.9]{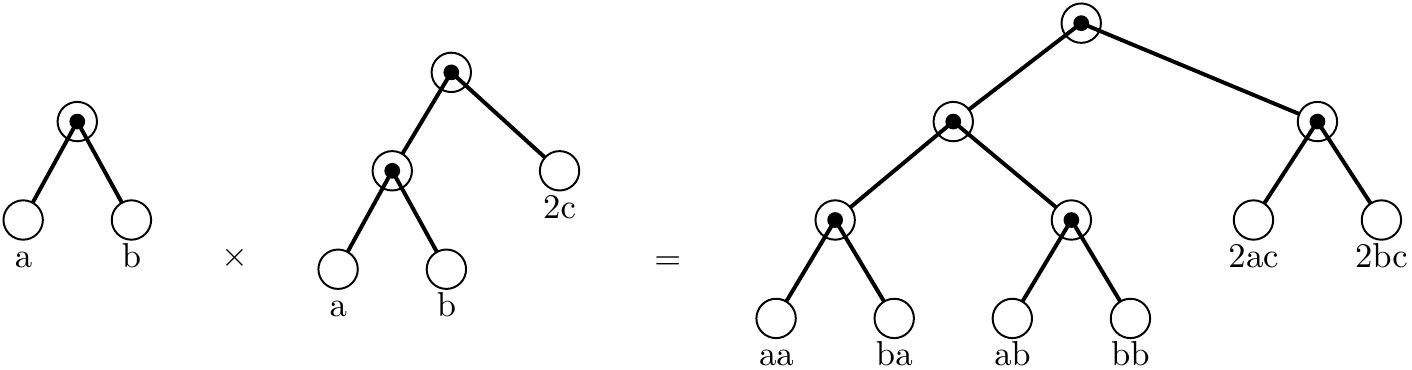}
\end{center}

\dodefn{5} For $X$ a leaf of a tree 
the $\Natural$ valued function {\it $X$ has depth $\cdot$ in $\cdot$} is
defined recursively as\\
(1) $X$ has depth $0$ in $\Tree(X).$ \\
(2) If $X$ has depth $k$ in $\Delta_Y$ then
$X$ has depth $k+1$ in any tree $\Tree(Y+Z, \Delta_Y, \Delta_Z).$
\tombstone
\vspace{5pt}

We also write this function as $\Depth(X,\Delta_Y),$
for $X$ a leaf of $\Delta_Y.$

\dodefn{6} 
The {\it weight} of a tree $\Delta_X$
is written $W(\Delta_X)\in\Natural$. 
This is the depth weighted sum of the norm of the leaves of $\Delta_X:$
$$
    W(\Delta_X) := \sum_{\Leaf(Y,\Delta_X)} \Depth(Y,\Delta_X) |Y|.
$$ \tombstone

First we show the weight function respects thickenings.
\dotheorem{7} 
For $k\in\Natural,$ tree $\Delta_X$ and multiset $Y$,
\begin{align*}
    W(k \cdot\Delta_X) &= k W(\Delta_X) \\
    W(Y \cdot\Delta_X) &= |Y| W(\Delta_X) 
\end{align*}
\doproof The left multiplication in Definition 2 \& 3 distributes over the leaves.
\tombstone


The next theorem says that $W(\cdot)$ is a derivation on trees.
\dotheorem{8} 
For trees $\Delta_X$ and $\Delta_Y$ we have
$$
    W(\Delta_X\times\Delta_Y) = |X|W(\Delta_Y) + W(\Delta_X)|Y|.
$$
\doproof By inspection of the construction of $\Delta_X\times\Delta_Y.$
\tombstone

Using this theorem we can write the cute formula:
$$
    W(\Delta_X^n) = n |X|^{n-1} W(\Delta_X).
$$

An important class of trees has only monomial multisets on leaves.
These trees correspond to fully parenthesized expressions for a multiset.
\dodefn{9} 
A {\it monomial tree over a multiset} is defined recursively as\\
(1) For any monomial $X$, $\Tree(X)$ is a monomial tree over $X$.\\
(2) If $X$ and $Y$ are disjoint multisets,
and $\Delta_X$ is a monomial tree over $X$ and
and $\Delta_Y$ is a monomial tree over $Y$, then
$\Tree(X+Y, \Delta_X, \Delta_Y)$ is a monomial tree over $X+Y.$
\tombstone

\dolemma{10} The product of two monomial trees is monomial. 
\doproof
By inspection of the product construction.
\tombstone

\dotheorem{11} 
For any monomial tree $\Delta_X$ we have
$$
    w(X) \le W(\Delta_X).
$$
\doproof This is well known, see~\cite{Huffman1952,Mackay2003} \tombstone

The next question is, when does equality obtain in this theorem?

%
%
%

\section{Huffman trees}



Given a multiset $X$,
The idea with Huffman coding is to
sequentially build a tree over $X$ from the bottom-up in an agglomerative way.
We start with a collection of trees, one for each monomial in $X$,
and then sequentially join pairs of trees, minimizing the norm
of the joined multisets as the sequence proceeds~\cite{Huffman1952}.

It turns out this algorithm is optimal, as opposed to
the top-down algorithm,
which is divisive: start with a single tree $\Tree(X)$ and sequentially
perform optimal splitting operations, such as $\Tree(X)\mapsto \Tree(X, \Tree(Y), \Tree(Z)).$

\dodefn{12} 
Given a multiset $X$ a {\it Huffman tree over $X$}
is any tree over $X$ that can be constructed as follows.
Writing $X=\sum n_a \cdot a\in\Natural[A]$
we proceed inductively:\\
(Step 1) Define a finite set of monomial trees 
$\Omega_1 := \{ \Tree(n_a\cdot a)\}_{a\in A}.$
\\
(Step $n+1$) Given the finite set of trees 
$\Omega_n = \{ \Delta_{X_1},...,\Delta_{X_m} \}$
we define $\Omega_{n+1}$ as follows.
Choose any $1\le i< j\le m$ such that
$$
    |X_i + X_j| \le |X_{i'} + X_{j'}| \ \ \forall 1\le i'< j' \le m.
$$
Then 
$$
    \Omega_{n+1} := \{\Delta_{X_1}, ... \widehat{\Delta_{X_i}},..., \widehat{\Delta_{X_j}}, ... \Delta_{X_m},
    \Tree(X_i+X_j, \Delta_{X_i}, \Delta_{X_j}) \}.
$$
(Final step.)
After a finite number of steps, $\Omega_{n}$ will contain
a single tree over $X$. This tree is the constructed Huffman tree.
\tombstone

It follows that any Huffman tree is a monomial tree.
Here we show a Huffman tree for the multiset $5a+5b+4c+3d+3e:$ 
\begin{center}
\includegraphics[scale=0.9]{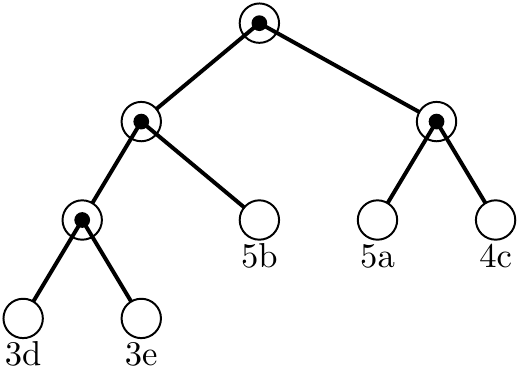}
\end{center}

We write $\Gamma_X$ for a generic Huffman tree over $X$.
The next theorem says that Huffman trees are 
characterized as those trees that minimize $W(\cdot).$

\dotheorem{13} 
For any monomial tree $\Delta_X$ we have
$ W(\Gamma_X) \le W(\Delta_X). $
Conversely, if $\Delta_X'$ is a monomial tree such that
$ W(\Delta_X') \le W(\Delta_X) $
for all monomial $\Delta_X$ then $\Delta_X'$ is a Huffman tree.
\doproof This is well known, see~\cite{Huffman1952,Mackay2003} \tombstone

\docorollary{14} 
If $\Gamma_X$ and $\Gamma_X'$ are two Huffman trees for the same
multiset $X$, then
$$
    W(\Gamma_X) = W(\Gamma_X').
$$
\tombstone

Because of this corollary, we may as well define $W(X) := W(\Gamma_X)$ where
$\Gamma_X$ is any Huffman tree for $X$.


\dodefn{15} 
A multiset $X\in\Natural[A]$ is {\it dyadic} when
$$
    X = \sum_{a\in A} 2^{k_a} \cdot a \ \ \mbox{and}\ \  \sum_{a\in A} 2^{k_a} = 2^k,
$$
with $k_a\in\Natural$ and $k\in\Natural.$
\tombstone

The next theorem says that decategorifying
Huffman trees gives the un-normalized entropy for
scalar multiples of dyadic multisets.

\dotheorem{16} 
$X$ is a scalar multiple of a dyadic multiset iff
$$w(X)=W(X).$$ 
\doproof See~\cite{Mackay2003} section 5.3. \tombstone

As a corollary, we obtain a categorification of Eq. (2):
\docorollary{17} 
For $k\in\Natural$ and $X$ a scalar multiple of a dyadic multiset, the tree
$$
    \Gamma_{kX} := k\Gamma_{X}
$$
is a Huffman tree for $kX.$
\tombstone

Here we show a Huffman tree on the dyadic multiset $a+b+2c+4d:$
\begin{center}
\includegraphics[scale=0.9]{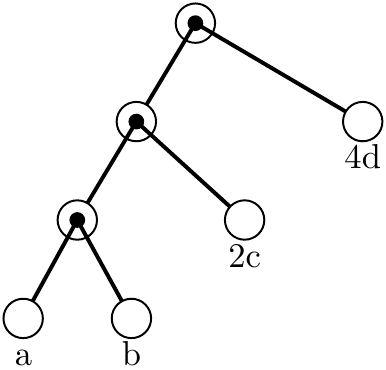}
\end{center}
This tree has weight $3+3+2\cdot 2+4 =14$.

The proof of the next theorem involves a mild categorification of Eq. (1).
\dotheorem{18} 
For disjoint dyadic multisets $X, Y$ with $X+Y$ dyadic, we have
$$
    W(\Gamma_{X+Y}) = W(\Tree(X+Y, \Gamma_X, \Tree(Y))) + W(\Gamma_Y).
$$
\doproof 
If $X, Y$ are disjoint dyadic multisets with $X+Y$ dyadic,
then $\Gamma_{X+Y} = \Tree(X+Y,\Gamma_X,\Gamma_Y)$ is a Huffman tree for $X+Y.$
\tombstone

The next theorem categorifies Eq.~(3).
It states that Huffman trees are a derivation over multisets.
\dotheorem{19} 
If the multisets $X$ and $Y$ are scalar multiples of dyadic multisets,
the product of Huffman trees $\Gamma_X$ and $\Gamma_Y$
is a Huffman tree for $X\times Y:$
$$
    \Gamma_{X\times Y} := \Gamma_X\times\Gamma_Y = X\Gamma_Y + \Gamma_X Y.
$$
\doproof 
Omitted.
\tombstone

\section{Final thoughts}


The importance of Huffman coding is highlighted by the
following limit, which holds for any multiset $X:$ 
$$
    \lim_{n\to \infty} \frac{W(X^n)}{n|X|^{n-1}} = w(X).
$$
Categorifying this limit seems beyond the techniques of the present work.




The recent independent work~\cite{Bradley2021} appears to be related.

{\bf Acknowledgments}
This work is inspired by a blog post of Tom Leinster~\cite{Leinster2019}
and a discussion with James Dolan.

\bibliography{refs_entropy}{}
\bibliographystyle{abbrv}

\end{document}